# The Great Division

**Yu Wang**
University of Rochester
Rochester, NY 14627
ywang176@ur.rochester.edu

**Jiebo Luo**
University of Rochester
Rochester, NY 14627
jluo@cs.rochester.edu

## Abstract

When information flow fails, when Democrats and Republicans do not talk to each other, when Israelis and Palestinians do not talk to each other, and when North Koreans and South Koreans do not talk to each other, mis-perceptions, biases and fake news arise. In this paper we present an in-depth study of political polarization and social division using Twitter data and Monte Carlo simulations. First, we study at the aggregate level people's inclination to retweet within their own ideological circle. Introducing the concept of *cocoon ratio*, we show that Donald Trump's followers are 2.56 more likely to retweet a fellow Trump follower than to retweet a Hillary Clinton follower. Second, going down to the individual level, we show that the tendency of retweeting exclusively within one's ideological circle is stronger for women than for men and that such tendency weakens as one's social capital grows. Third, we use a one-dimensional Ising model to simulate how a society with high cocoon ratios could end up becoming completely divided. We conclude with a discussion of our findings with respect to fake news.

## Introduction

*"One of the dangers of the Internet is that people can have entirely different realities, they can be just cocooned in information that reinforces their current biases."*
— former President Barack Obama

We live in the age of social media where Facebook, Twitter and Snapchat, to name just a few, constitute an integral part of our life. One of the distinctive features of social media is that we can easily find and befriend people with no more than a few clicks (Wang, Zhang, and Luo 2017). As a result, we can easily connect with hundreds if not thousands of people (Kwak et al. 2010) with little geographical restriction (Tranos and Nijkamp 2013).

The easiness of finding friends, in turn, is giving us the luxury of clustering only with like-minded people (McPherson, Smith-Lovin, and Cook 2001; Barberá 2015). It is part of our human nature to connect with people similar to us and avoid individuals who are very different, because distance is costly (Katelyn Y. A. McKenna 2000). Such distance can be, for example, geography (Kwak et al. 2010), interests (Bisgin, Agarwal, and Xu 2010) or political views (Wang, Zhang, and Luo 2017).

The easiness of connecting with people on social media and our inclination to cluster around similar people, unfortunately, combine to sow the perfect seed for social division and polarization. People of shared opinions, such as Tea Party[1], Vote Leave[2] and Hillary for Prison[3], group together, circulate, and reinforce their shared ideas via, for example, retweets and @mentions. The result is bias, mis-perception, and fake news. As former president Barack Obama warns, "one of the dangers of the Internet is that people can have entirely different realities, they can be just cocooned in information that reinforces their current biases."

In this paper, we present an in-depth analysis of this phenomenon using Twitter data and Monte Carlo simulations. In particular, we focus on the retweeting behavior. We first study at the aggregate level people's inclination to retweet within their own ideological circle. Introducing the concept of *cocoon ratio*, we show, for example, that Donald Trump's followers are 2.56 more likely to retweet a fellow Trump follower than to retweet a Hillary Clinton follower. Second, going down to the individual level, we show that the tendency of retweeting within one's own ideological circle is stronger for women than for men and that such tendency gradually weakens as one's social capital grows. Third, we use a one-dimensional Ising model to simulate how a society with a high *cocoon ratio* might end up being closed and isolated.

We summarize our contributions as follows:

- In a time of fake news and political polarization, we provide a thorough analysis of social cocooning on the internet.
- Going down to the individual level, we analyze how social capital and gender might play a role in the development of social division.
- We provide Monte Carlo simulations to better understand how our society might veer toward complete social division.
- We make our data and code publicly available to facilitate replication and further research.

---

[1] https://twitter.com/teapartyorg.
[2] https://twitter.com/vote_leave.
[3] https://twitter.com/HRC4Prison.

# Related Literature

Our work draws inspiration from branches in political science, sociology and social media studies. We divide the related literature into three categories: homophily in sociology and social media studies, polarization in American politics (and to some extent fake news), and studies of the retweeting behavior.

## Homophily, the Internet, and Social Fragmentation

Social clustering, which is the opposite side of polarization, represents the idea that birds of the same feather flock together (McPherson, Smith-Lovin, and Cook 2001; Barberá 2015; Goodreau, Kitts, and Morris 2009). People tend to cluster along lines of socio-demographic and interpersonal characteristics such as race, marriage, friendship, gender, religion, age, class, and education (Nahon and Hemsley 2014; Wang, Zhang, and Luo 2017).

Regarding the political effects of the internet, some scholars have argued that the emergence of internet-mediated forms of communication increases public participation and enhance individual freedom, cultural diversity, political discourse, and justice (Woodly 134; Benkler 2007). Some others, however, have warned against the possibility of people using "the Internet to listen and speak only to the likeminded" (Sunstein 2009).

Several earlier studies of political blogs, for example, have found that liberals tend to link to other liberal blogs and conservatives tend to link to other conservative blogs (Adamic and Glance 2005; Nahon and Hemsley 2014). Building on this line of research, our paper attempts to detect and analyze homophily in individuals' retweeting behavior.

## Polarization in American Politics

Earlier works have studied the increasing polarization of American politics at both the elite level (Hare and Poole 2014; McCarty, Poole, and Rosenthal 2009) and the mass level (Campbell 2016; Doherty 2014; Fiorina and Abrams 2008; Jacobson 2016). Druckman *et al.*, in particular, study how elite partisan polarization affects public opinion formation and find that party polarization decreases the impact of substantive information (Druckman, Peterson, and Slothuus 2013).

The phenomenon of polarization became all the more apparent during the 2016 presidential election, where Trump and Clinton depicted two radically different pictures of America. Trump tried to paint "a picture of a bitter, broken country," whereas Hillary Clinton would like to highlight "the energy and optimism" in America (Clinton 2017). From a language and visual perspective, Wang et al. (Wang et al. 2017) studies to what extent Trump followers on Twitter can be distinguished from Hillary Clinton's followers.

## Retweeting

Retweeting, most commonly represented with the form "RT @user message," is Twitter's equivalent of email forwarding (Boyd, Golder, and Lotan 2010). It serves as a form of information dissemination (Kwak et al. 2010), a means of participating in diffuse conversations and a way to publicly agree with someone (Boyd, Golder, and Lotan 2010). Empirical studies have found that people are more likely to retweet messages from friends (Boyd, Golder, and Lotan 2010) and ones that they are interested in (Luo et al. 2013; Lee et al. 2015; Yang et al. 2010; Cha et al. 2010).

A consistent finding among these studies is that people tend to retweet messages of interest and messages from friends. Since Trump (Clinton) followers share the same interest in Trump (Clinton) and by the principle of homophily they are more likely to be friends in the first place, we hypothesize that Trump followers are more likely to retweet fellow Trump followers than to retweet Clinton followers and that Clinton followers are more likely to retweet fellow Clinton followers than to retweet Trump followers.

# Data

## Defining Candidate Followers

The core of our dataset is the tweets posted by Trump followers and Hillary followers. Here, we set our target users as the 7.497 million people who were following Donald Trump on Twitter on April 9th, 2016 and from the 5.905 million people who were following Hillary Clinton on April 10th, 2016. We choose these pre-election days in April so that the action of following could serve as a stronger signal of interest and support as compared to after the election in November 2016. Throughout we do not assume that all followers are supporters.

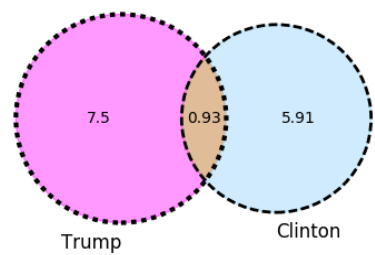

Figure 1: Of the 12.48 million Trump followers and Hillary followers, only 0.93 million people are following both of them. In our experiments, we select data exclusively from people who only follow Trump (pink) and who only follow Clinton (blue). Donald Trump and Hillary Clinton are not included in these samples.

Following (Wang et al. 2017), we remove individuals who were following both Trump and Clinton, which amount to 930,783 people (Figure1). Then we randomly sample 1 million users among Trump followers (purple) and 1 million users among Clinton followers (blue). For each selected in-dividual, we collect up to 100 tweets. Data collection was carried out in December 2017.

**100 Tweets Per Person**

We set the limit of data collection to 100 tweets per person rather than the 3,000 maximum allowed on Twitter so as to alleviate the problem of unbalanced representation, as an individual with 3,000 tweets could have a representation equivalent to 3,000 individuals with only one tweet. We do not follow the 1 tweet per person principle either, because we do need substantial numbers of tweets to capture an individual's (re)tweeting pattern. As a result, we set the limit to 100 as a compromise.

Of the 1 million Hillary followers, we find that 130,885 (13%) accounts are either private or no longer exist, as a result of which we do not collect postings from them.[4] From the remaining 869,115 public accounts, we collected 41,174,942 tweets from Clinton followers. Of the 1 million Trump followers, we find that 141,723 (14%) accounts are either private or have been deleted. From the remaining 858,277 Trump followers, we were able to to collect 41,779,547 tweets.

**Data Attributes**

For each tweet, we store the message, the creation date, the id of the message and the user id. Besides tweet attributes, for each individual we also collect their screen name, for example *@PhilippeReines*, name, for example *Philippe Reines*, and the number of followers. We use screen name to uniquely identify individuals as they are used as handlers in retweets and mentions. We infer gender from a person's name (Wang, Zhang, and Luo 2017). We use the number of followers as a proxy for social capital (Wang, Li, and Luo 2016).

**Data Distribution**

In Table 1, we report the percentile distribution of the number of tweets per person. It can be seen that for over 30 percent of Clinton's followers and 20 percent of Trump followers in our sample set, we collected 100 tweets. For the bottom 10% of the sampled individuals, however, we only collected 2 tweets. This holds true for both Hillary Clinton and Donald Trump.

Table 1: Distribution of the Number of Tweets per Individual

| Percentile | Hillary Clinton | Donald Trump |
|---|---|---|
| 90% | 100 | 100 |
| 80% | 100 | 100 |
| 70% | 100 | 99 |
| 50% | 70 | 73 |
| 40% | 70 | 73 |
| 30% | 28 | 30 |
| 20% | 5 | 5 |
| 10% | 2 | 2 |

---

[4]The list of such accounts is reported in the first author's website for verification and reproduction.

In Figure 2, we report the time-series distribution of our data. The y axis is the natural logarithm of the number of tweets posted and the x axis is each day from 2008 to 2017. It can be seen that the oldest tweets in our collection date back to 2008. A straight line fits the data points reasonably well, suggesting that the number of tweets in our collection grew exponentially with most tweets falling into the year of 2017.

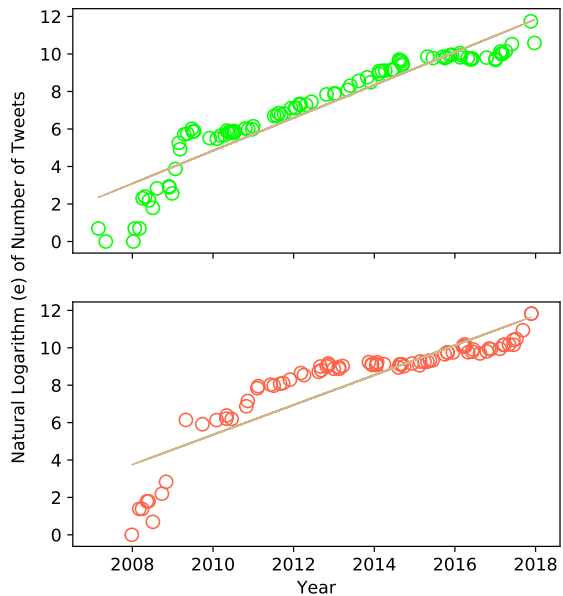

Figure 2: The time-series distribution of tweets from Hillary Clinton's followers (Upper) and from Donald Trump's followers (Lower). For better visualization, we randomly sampled 2% of the original data points.

**Processing Retweets**

From each tweet, we extract two main signals: *retweet* and *mention*. The first example below is a retweet and it represents an edge between the retweeter and the originator. The second tweet mentions @glennbeck. Twitter mentions represent another directed edge between originator and the person being mentioned. In our discussion, we will mostly focus on retweets and only briefly report on the summary statistics of the mentions.

*RT @OutVeo: There's nothing quite like having experts build your business for you is there?*

*@glennbeck He does have tendency to get off subject and ramble as if we wanted to hear that part.*

Using regular expression, we extract all the Twitter users who are being retweeted or mentioned. Regarding recursive retweets, i.e., retweets of the form RT @A RT @B msg (Yang et al. 2010), we treat user A as the originator of the tweet and discard B.

# Empirical Findings

In this section, we present our main findings. First, we present aggregate analysis of the retweet patterns of Trump followers and Clinton followers. Second, we perform econometric analysis at the individual level and model an individual's openness to opposite opinions as a function of gender, social capital and ideology. Third, we simulate the dynamics of social openness using an Ising model.

## Information Cocoon

Our study is inspired by President Obama's warning against the dangers of individual citizens getting 'cocooned' in bias via social media and by the prevalence of fake news stories (Allcott and Gentzkow 2017a; Jin et al. 2017b) surrounding the 2016 presidential election. In this subsection, we show that Trump (Clinton) followers are much more likely to retweet a fellow Trump (Clinton) follower.

Table 2: Retweet Matrix

| Source | Clinton | Trump | Other | Cocoon Ratio |
|---|---|---|---|---|
| Clinton | 2,028,960 | 497,742 | 13,321,216 | 4.08 |
| Trump | 547,952 | 1,405,751 | 12,517,074 | 2.56 |

**The Retweet Matrix** In total, we found that 15,847,918 tweets out of the 41,174,942 tweets (38.5%) by Clinton followers are retweets. Among the 15.8 million retweets, 2,028,960 originated from other Clinton followers. By contrast, only 497,742 retweets originated from Trump's followers. We found 14,470,777 tweets out of the 41,779,547 tweets by Trump followers are retweets (34.6%). Among the 14.5 million retweets, 1,405,751 tweets originated from other Trump followers. Only 547,952 tweets originated from Clinton followers.

Here we introduce the idea of *cocoon ratio*. We define *cocoon ratio* as the ratio of retweeting one's own candidate's followers to retweeting the opposite candidate's followers. For example, cocoon ratio for Hillary Clinton's followers is calculated as follows:

$$\text{Cocoon Ratio}_c = \frac{\#\{\text{retweets of Clinton followers}\}}{\#\{\text{retweets of Trump followers}\}}$$
$$= \frac{2,028,960}{497,742}$$
$$= 4.08$$

We observe that a Clinton follower is 4.08 times more likely to retweet a fellow Clinton follower than a Trump follower and that a Trump follower is 2.56 times more likely retweet a fellow Trump follower than a Clinton follower.

**The Mention Matrix** In a similar vein, we calculate the cocoon ratio based on Twitter mentions (Table 3). Consistent with retweets, we find that Trump (Clinton) followers are more likely to mention other Trump (Clinton) followers in a tweet, thus reinforcing Trump (Clinton) followers as a group.

Table 3: Mention Matrix

| Source | Clinton | Trump | Other | Cocoon Ratio |
|---|---|---|---|---|
| Clinton | 2,137,774 | 712,437 | 18,532,273 | 3.00 |
| Trump | 910,439 | 2,216,796 | 19,488,038 | 2.43 |

The root cause of these high cocoon ratios, we suspect, lies in the fact that a person's friends usually follow the same candidate as the person does. It is psychologically and ideologically less costly to befriend people who are similar to oneself than to befriend people who are different (McPherson, Smith-Lovin, and Cook 2001; Barberá 2015). Political ideology, in turn, constitutes an integral component of that similarity function.

## Time Series Analysis

*"To all Republicans and Democrats and independents across this nation, I say it is time for us to come together as one united people."*

– President Donald Trump

During his victory speech on November 9th, 2016 after winning the U.S. election, President-elect Donald Trump called on people of all ideologies to heal the division and unite under the flag. One important question to ask is whether or not people have responded to the president's call.

In this subsection we analyze how the cocoon ratio has been evolving before and after the election on a monthly basis between 2015, in which both Hillary Clinton and Donald Trump declared their presidential candidacy, and 2017, which is one year into Trump's presidency.

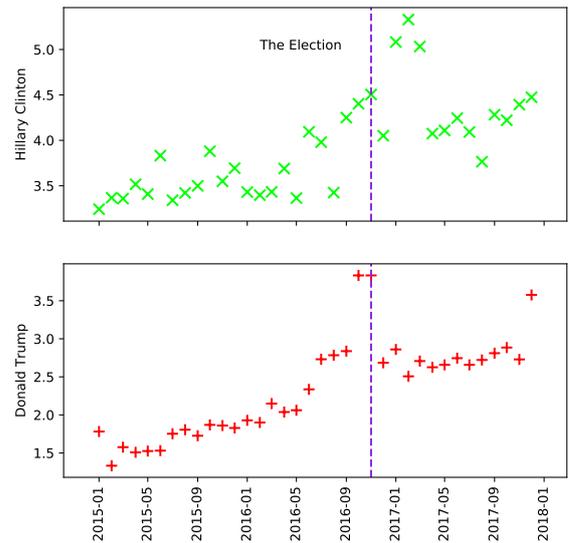

Figure 3: Monthly cocoon ratio calculated for Hillary Clinton's followers (upper) and Donald Trump's followers (lower).

We group the retweets based on the month and year in which the tweets were posted and calculate the cocoon ratio for each month separately. We report our findings in Figure 3, where each data point represents a month. We find that the cocoon ratio has been increasing for both Hillary Clinton and Donald Trump throughout the presidential campaign. The increase only accelerated in the last months of the election and peaked in November.

We also notice that after the election the cocoon ratio has been staying at a high level rather than return to the pre-election levels. This suggests that union between the Democrats, the Republicans and the Independents has not realized, contrary to what President Trump called for in his victory speech.

**Econometric Modeling**

In this section, we model the cocoon phenomenon using individual attributes: gender, social capital and ideology. First, we only select individuals from whom we have collected 100 tweets. As reported in Table 1, this amounts to more than 0.4 million samples. For each of those selected individuals, we code as one if they have retweeted a follower of the opposite candidate (e.g. a Clinton follower retweeting one of Donald Trump's followers) and zero otherwise. Mathematically, if the probability of a person retweeting someone who follows the opposite candidate is p, assuming i.i.d, then the probability of retweeting that person within 100 tweets is $1-(1-p)^{100}$. We refer to this variable as *Open*, indicating "open-minded." Figure 4 illustrates the relationship between the probability of retweeting the opposite candidate and the probability of *Open*=1.

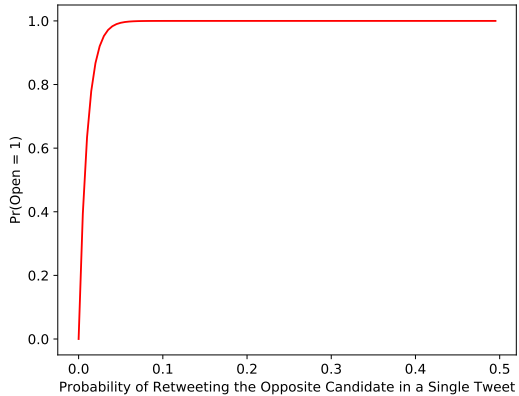

Figure 4: The probability of *Open*=1 is larger than 95% when the probability of retweeting followers of the opposite candidate is greater than 3%.

We infer gender information from a person's first name (Wang, Zhang, and Luo 2017; Mislove et al. 2011). For this purpose, we have compiled a list of 800 names, based on their appearance frequencies on Twitter, that are gender-revealing, such as Mike, David, Emily and Sarah.[5]. We use the number of followers as a proxy for social capital (Wang, Li, and Luo 2016), and define *social capital* as log(1+#followers). We include both its linear and quadratic terms in the regression.

We report the summary statistics in Table 4. For 56% of our observations, *Open* = 1. 45% of the observations are women. 54% are following Hillary Clinton. The mean of *social capital* is 5.33.

Table 4: Regression Summary Statistics

| Variable | # Observations | Mean | Std. Dev. | Min | Max |
|---|---|---|---|---|---|
| Open | 379,676 | 0.56 | 0.50 | 0 | 1 |
| Female | 171,703 | 0.45 | 0.50 | 0 | 1 |
| Hillary | 379,676 | 0.54 | 0.50 | 0 | 1 |
| Social Capital | 379,676 | 5.33 | 1.52 | 0 | 17.80 |
| Social Capital$^2$ | 379,676 | 30.77 | 18.06 | 0 | 316.72 |

Table 5: Logistic Regression

| | 1 | 2 | 3 |
|---|---|---|---|
| **Open** | | | |
| Social Capital | 0.0429*** | 0.0482*** | 0.219*** |
| | (0.000) | (0.000) | (0.000) |
| Hillary Clinton | -0.602*** | -0.572*** | -0.573*** |
| | (0.000) | (0.000) | (0.000) |
| Female | | -0.106*** | -0.106*** |
| | | (0.000) | (0.000) |
| Social Capital$^2$ | | | -0.0151*** |
| | | | (0.000) |
| Constant | 0.324*** | 0.490*** | 0.0435 |
| | (0.000) | (0.000) | (0.286) |
| Observations | 379459 | 171633 | 171633 |
| Pseudo $R^2$ | 0.016 | 0.017 | 0.017 |

*p*-values in parentheses
* p¡0.10, ** p¡0.05, *** p¡0.010

We report our results in Table 5. Consistent across all three specifications, we find that individuals with higher social capital (i.e. more followers) are more likely to retweet followers of the opposite candidate. This suggests that people with large personal networks are more likely to be talking with people of different opinions than people with only a small personal network.

We also find that followers of Hillary Clinton are less likely to retweet Trump followers than Trump followers are to retweet Clinton followers. Lastly, compared with men, women are less likely to retweet across the party line, consistent with previous findings (Wang, Zhang, and Luo 2017; McPherson, Smith-Lovin, and Cook 2001; Wang et al. 2016a; 2016b).

---

[5] The complete list is available on the first author's website.

## Modeling Dynamics of Democratic Opinions

In this subsection, we study how the variable *open* for Trump's and Clinton's followers evolves using a one-dimensional Ising model (Sznajd-Weron and Sznajd 2000; Holme and Newman 2006; Stauffer 2010). As reported above, around 56% of our observations are marked $Open$, i.e., they retweet followers of the opposite candidate. Here we mark the remaining 44% as $Closed$. We want to explore the steady state of the mixture of $Open$ individuals and $Closed$ individuals: is it closed or open or a mixture of the two?

Here we consider an Ising spin chain (S$i$; i = 0, 1, . . . N-1). Each spin is either $Open$ or $Closed$. We denote $Open$ with 1 and $Closed$ with -1. The dynamic rules are as follows:

- if $S_iS_{i+1}$ = 1 then $S_{i1}$ and $S_{i+2}$ take the direction of the pair $S_i$, $S_{i+1}$, for i = 2, 3, ... , N-1.
- if $S_iS_{i+1}$ = -1 then $S_{i1}$ takes the direction of $S_{i+1}$ and $S_{i+2}$ takes the direction of $S_i$, for i = 2, 3, ... , N-1.

For example, if an individual (i-1)'s two neighbors (marked as i, i+1) are both closed, then individual i will become the same as the two neighbors. If one neighbor is closed, and the other open, then individual i will take the same value as individual i+2. In our experiments, i takes values between 1 and N-3, inclusive.

Following the same spirit of stochastic gradient descent, each time we randomly choose i and update $S_{i-1}$ and $S_{i+2}$ according to the above dynamic rules (Krapivsky and Redner 2003; Liggett 1985).

Here we set N=200, and initialize the spins with a random number generator. Each spin will be set $Open$ with a probability of 0.53, the same as our empirical observation. With probability of 0.47, we set them to $Closed$.

There are three steady states of this model:

- *Closed:* all individuals become $Closed$, i.e., they only retweet people following the same candidate.
- *Open:* all individuals become $Open$, i.e., everyone is open to retweeting people following the opposite candidate.
- *Mixed:* 50% of the individuals become $Closed$ and the other 50% become $Open$. For example, one possible permutation could be [Closed, Open, Closed, Open, Closed, Open, Closed, Open, Closed, Open, Closed, Open, Closed, Open, **...** ]

We define the overall $Climate$ as the mean of all individuals' orientations:

$$Climate = \frac{1}{N} \sum_{i=1}^{i=N} S_i$$

In Figure 5, we illustrate one realization of the dynamics of $Climate$ with 1 million iterations (Sznajd-Weron and Sznajd 2000). For each spin, we set it to 1 (denoting $Open$) with probability 0.53, as observed in our empirical data. It can be seen that $Climate$ is approximately 0.06 at the initialization. It increases to as high as 0.9 but keeps fluctuating. After 0.4 million iterations, the mixed steady state of [..., $Closed$, $Open$, $Closed$, $Open$, $Closed$, $Open$, $Closed$, $Open$...] is achieved.

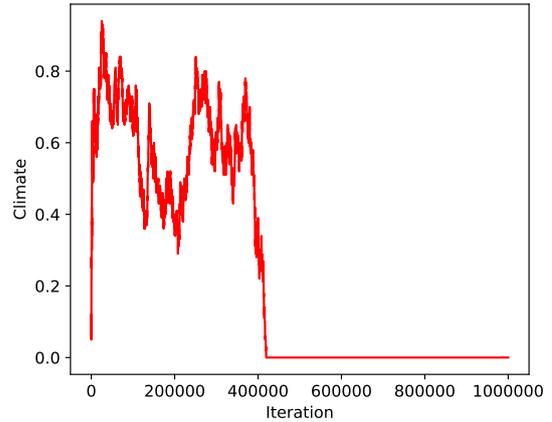

Figure 5: At initialization, $Climate$ is around 0.06, then it keeps fluctuating. It reaches the mixed steady state after 400,000 iterations.

Next, we vary the percentage of $Closed$ spins to analyze its effects on the distribution of the steady state. We summarize our findings in Figure 6. As the number of $Closed$ spins increases, the probability of the society taking the closed steady state [... $Closed$, $Closed$, $Closed$, $Closed$, $Closed$, $Closed$, ...] increases. The probability of reaching the mixed steady state first increases, reaches its peak when half of the spins are $Closed$ at initialization, and then decreases towards zero.

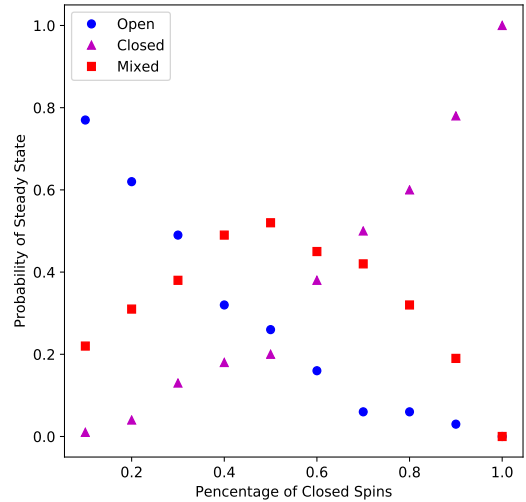

Figure 6: As the percentage of $Closed$ increases from 0 to 1, the probability of the society ending up as $Closed$ grows to 1.

## Discussion with Respect to Fake News

*"The difficulty of unmasking and eliminating fake news is due also to the fact that many people interact in homogeneous digital environments impervious to differing perspectives and opinions. Disinformation thus thrives on the absence of healthy confrontation with other sources of information that could effectively challenge prejudices and generate constructive dialogue; instead, it risks turning people into unwilling accomplices in spreading biased and baseless ideas."*

– Pope Francis

The 2016 U.S. presidential election has witnessed the rise of fake news and social bots (Allcott and Gentzkow 2017b; Bessi and Ferrara 2016; Qiu et al. 2017). It is reported that the most viral fake news stories were more widely shared on Facebook than the most popular mainstream news stories (Silverman 2016). Some commentators even suggest that fake news has affected the U.S. election result (Dewey 2016; Read 2016).

Faced with the threat of fake news, researchers have experimented with various algorithms in an effort to identify fake news, including SVM, decision trees, decision rules, Bayes networks (Castillo, Mendoza, and Poblete 2011; Jin et al. 2017b), recurrent neural network (Ma et al. 2016). Researchers have explored both texts (Horne and Adal 2017) and images (Jin et al. 2017a) as data sources.

Such research certainly plays an important role in reducing the amount of fake news. We suggest, however, that another effective way to counter fake news is to help people talk to each other (Sunstein 2009). This is because when people holding different perceptions (Jervis 1976), ideas and political views come together and openly discuss and debate with each other, instead of cocooning in their own block, fake news naturally disappears.

When information flows freely across communities, when Democrats and Republicans talk to each other, when Israelis and Palestinians talk to each other, and when North Koreans and South Koreans talk to each other, mis-perceptions, biases and fake news fade away.

## Conclusion

In this paper we presented an in-depth study of political polarization and social division using social media data and Monte Carlo simulations. First, we studied at the aggregate level people's inclination to retweet within their own ideological circles. We introduced the concept of *cocoon ratio* and showed that Donald Trump's followers are 2.56 more likely to retweet a fellow Trump follower than to retweet a Hillary Clinton follower. Second, going down to the individual level, we showed that the tendency of retweeting exclusively within one's own ideological circle is actually stronger for women than for men and that such tendency gradually weakens as one's social capital grows. Third, we used a one-dimensional Ising model to simulate how a society with high cocoon ratios might end up becoming a completely *closed* society. We discussed our findings with regards to fighting fake news and suggested that encouraging open-minded discussion can serve as an effective way to counter fake news.